\documentclass[aps,prc,twocolumn,twoside,showkeys,showpacs,floatfix]{revtex4}
\usepackage{graphics,epsfig}
\graphicspath{{./},{./EPS2/}}

\newcommand{\figart}[5]{\begin{figure}[#1]
			\begin{center}
                        \includegraphics[scale=#5,draft=false]{#2.eps}
                        \caption{\small \it #3}
                        \label{#4}
                        \end{center}
                        \end{figure}}
			
\newcommand{\figartlarg}[5]{\begin{figure*}[#1]
			\begin{center}
                        \includegraphics[scale=#5,draft=false]{#2.eps}
                        \caption{\small \it #3}
                        \label{#4}
                        \end{center}
                        \end{figure*}}
			
\newcommand{\LPC}[0]{\it LPC Caen (IN2P3-CNRS/ISMRA et Universit\'e),
14050 Caen Cedex , France}

\newcommand{\Elb}[0]{$E_{LeastBound}$}
\newcommand{\Ebind}[0]{$E_{bind}/N$}
\newcommand{\Ecm}[0]{$E_{c.m.}/N$}
\newcommand{\bred}[0]{$b_{red}$}
\newcommand{\Estar}[0]{$E^*/N$}
\newcommand{\Np}[0]{$N_{proj}$}
\newcommand{\Nt}[0]{$N_{targ}$}
\newcommand{\Vpar}[0]{$V_{//}$}

\newcommand{\mod}[1]{{#1}}

\begin{document}

\title{\Large \bf Limitation of energy deposition in Classical N Body
Dynamics.}
\author{D.~Cussol}
\affiliation{\LPC}
\date{\today}

\begin{abstract}
Energy transfers in collisions between classical clusters 
are studied with Classical N Body
Dynamics calculations for different entrance channels. 
It is shown that the energy per particle transferred to \mod{thermalised} 
classical
clusters does not exceed the energy of the least bound particle in the cluster
\mod{in its ``ground state''}.
This limitation is observed during the whole time of the collision, except for
the heaviest system. 
\end{abstract}

\maketitle


The question of energy deposition in nuclei during 
nucleus-nucleus collisions is
of great importance for nuclear matter studies. 
The maximum amount of energy which
can be stored in \mod{hot equilibrated} nuclei has been studied both 
experimentally 
\cite{Jiang89,Vient94,Nato2002,Schmidt2002,Galin94,Steck2000} and theoretically 
\cite{BonLevVaut84,BonLevVaut86}. Such studies
have mainly been motivated by the determination of a plateau in the so called 
caloric curve
(the evolution of the temperature with the excitation energy) which could be a
signature of a first order liquid-gas phase transition
\cite{Ma98,Poch95}. 

Experimentally, a lot of work has been done to determine the amount of 
\mod{thermal} energy stored
in nuclei. For central collisions, large energy deposits up to
22.5 A.MeV have
been determined \cite{Borderie96,Ma98}.
Other studies have shown that the excitation energy in primary products in
multifragmenting systems is around 3-4 A.MeV
\cite{Schmidt2002,Hudan,Marie98,Nato2002}, far below the total
available energy in the center of mass frame. This energy does not seem to
evolve strongly with the incident energy. These two measurements seem to be
in contradiction.
Possible limitations of energy deposition could result from 
prompt emission of energetic light charged particles at  
early times in the reaction \cite{Lefort2000,Dore2000a,Dore2000b}.

Theoretically, the maximum energy that \mod{a equilibrated} 
nucleus can withstand 
corresponds to the energy (or temperature) at which the surface
tension vanishes. This is often characterised by a critical temperature
$T_{c}$ whose value is around 16 MeV \cite{BonLevVaut84,BonLevVaut86}. 
This temperature is linked to the equation
of state of nuclear matter. The main drawback of such studies is that \mod{they
assume that the system is fully equilibrated and hence} do
not take into account possible limitations coming from the reaction mechanism.

The aim of the present article is to study energy deposition during collisions
between finite size systems in a well
controlled framework. Results from the Classical N Body Dynamics code
\cite{CNBD2002} will be shown
and the mechanism of energy deposition in classical clusters will be studied.
\mod{We will be interested in this paper by the energy transfered to 
thermally equilibrated clusters which corresponds to the energy that
long-lived clusters can withstand.} 
The
paper is organised as follows. In a first section, the Classical N Body
Dynamics will be briefly described. The excitation energy in clusters will 
be
shown for various systems and incident energies in the second section. The third
section will be devoted to the mechanism of energy deposition in clusters. 
Conclusions will be drawn in the last section 
 
\begin{table*}[ht]
\caption{Summary of systems. \label{t:systemes}}
\begin{ruledtabular}
\begin{tabular*}{0.5cm}{c c c c c c c}
~~$N_{projectile}$~~&~~\Nt~~&~~\Ecm~~&~~\Ebind fused~~& 
~~\Elb fused~~&~~\Estar~fused~~&~~number of events~~\\
~~&~~&~(E.S.U.)~&~(E.S.U.)~& 
~(E.S.U.)~&~(E.S.U.)~&~\\
\hline
\hline
13 & 13 & 25 & -73.30 & -50.86 & 35.17 & 1,000 \\
&  & 45 & & &55.17 &1,000 \\
&  & 65 & & &75.17 &1,000 \\
&  & 85 & & &85.17 &1,000 \\
\hline
18 & 50 & 30 &  -93.60 & -54.64 & 41.55 &1,000 \\
 &  & 60 &  & &71.55 &1,000 \\
 &  & 90 &  & &101.55 &1,000 \\
 &  & 120 &  & &131.55 &1,000 \\
\hline
34 & 34 & 30 & -93.60 & -54.64 & 42.04 &1,000 \\
 &  & 60 & & &72.04 &1,000 \\
 &  & 90 & & &102.04 &1,000 \\
 &  & 120 & & &132.04 &1,000 \\
\hline
50 & 50 & 30 & -98.93 & -62.46 & 41.50 &1,000 \\
 &  & 60 & & &71.50 &1,000 \\
 &  & 90 & & &101.50 &1,000 \\
 &  & 120 & & &131.50 &1,000 \\
\hline
100 & 100 & 30 & -107.97 (*) & -65.00 (*) & 39.03 (*)& 100 \\
 &  & 60 & & &69.03 (*)&100 \\
 &  & 90 & & &99.03 (*)&100 \\
 &  & 120 & & &129.03 (*)&100 \\
\end{tabular*}
\end{ruledtabular}
\end{table*}

\section{Description of the code.}
Let us start by describing the Classical N-Body Dynamics code (labeled CNBD)
used in this article.
The basic ingredients of such a code are very simple. The dynamical evolution of 
each particle of
the system is driven by the classical Newtonian equations of motion. The
two-body potential used in the present work is a third degree polynomial whose
derivatives are null at the range $r_1$ and at the distance of maximum depth
$r_{min}$. The depth value is $V_{min}$ and the value at $r=0$ is finite and equal 
to $V_0$. This potential has the basic properties of the Lennard-Jones
potential used in other works \cite{Dorso99,Sator2001}: 
a finite range attractive part and a repulsive
short range part. To follow the dynamical evolution of the system 
an adaptative stepsize fourth-order Runge-Kutta algorithm is used 
\cite{NumRec}. The main difference with other works is that the time step
$\Delta t$ can vary: if the potential varies strongly, $\Delta t$ is small and
when the potential varies gently, $\Delta t$ becomes larger. 
This allows a very high
accuracy with shorter CPU time than for fixed time step algorithms.
It requires an additional simulation parameter $\epsilon$ which is adjusted 
to ensure the
verification of conservation laws (energy, momentum, angular momentum) with a
reasonable simulation time. The energy difference between the beginning and the
ending simulation time is lower than 0.001\%.
This simulation has five free parameters: four
linked to the physics (the interaction) and one linked to the numerical
algorithm.

Since one wants to study the simplest case, neither long range repulsive
interaction nor quantum corrections like a Pauli potential have been introduced
\cite{Dorso87}.
Additionally, no statistical decay code is applied to the excited fragments 
formed during the collision. The final products have to be regarded as
``primary'' products which would decay afterwards. 
\mod{Although most of the ingredients necessary for a correct description of
atomic nuclei are missing from this simulation, it should be noted that
the reaction mechanisms
observed in nucleus-nucleus collisions are seen,}
and the properties of the ``ground states'' of such clusters are qualitatively
close to those of nuclei \cite{CNBD2002}.

In order to avoid any confusion with nuclear physics, the units used here are
arbitrary and called Simulation Units ($S.U.$).  The distance will then be in
Distance Simulation Units ($D.S.U.$), the energies in Energy Simulation Units
($E.S.U.$), the velocities in Velocity Simulation Units ($D.S.U./T.S.U.$) and
the reaction time in Time Simulation Units ($T.S.U.$). 

For this study, 16,400 events have been generated. For a fixed projectile size
\Np, a
fixed target size \Nt~and a fixed available energy in the center of 
mass \Ecm, the impact 
parameter $b$ is randomly chosen assuming a flat distribution between 0 and
$b_{max}=r(N_{proj})+r(N_{target})+r_1$, where $r(N_{proj})$ is the mean square
radius of the projectile, $r(N_{target})$ is the mean square
radius of the target and $r_1$ the range of the potential. In the analyses, each
event is weighted by its impact parameter value ($weight \propto b$). 
The systems and the
energies studied here are summarized in table \ref{t:systemes}. As it can be seen
in table \ref{t:systemes}, the available energies in the center of mass frame
\Ecm~have been chosen in such a way that the excitation energy of the fused
cluster \Estar~is close to the binding energy of the fused
cluster ($E_{c.m.}/N = 90\> E.S.U.$), close to the energy of the least bound particle
in the fused cluster ($E_{c.m.}/N = 60\> E.S.U.$), far below 
($E_{c.m.}/N = 30\> E.S.U.$) or far above ($E_{c.m.}/N = 120 \>E.S.U.$) the binding
energy of the fused cluster. The stars for the 100 + 100 system mean that these
energies were estimated from the liquid-drop parametrisation of the binding
energies of the clusters \cite{CNBD2002}. In that case, the energy of the least
bound particle is taken equal to $-65 \> E.S.U.$. 
For each system, the excitation energy $E^*_{fused}$ of
the fused cluster is determined by:

\begin{equation}
E^*_{fused} = E_{c.m.} + E_{bind}(N_{proj}) +E_{bind}(N_{targ})-
E_{bind}(N_{fused}) 
\end{equation}

where $E_{bind}(N)$ is the binding energy of the cluster of size $N$.

\mod{In this unit system, the parameters of the reaction are the following:
$V_0= 540 \>E.S.U.$, $V_{min}=-20 \>E.S.U.$, $r_{min}=10\>D.S.U.$ and
$r_{1}=15\>D.S.U.$. Typical mean square radii are around 10 $D.S.U.$ for 
$N\approx20$ and around 15 $D.S.U.$ for $N\approx70$ (for more 
information, see \cite{CNBD2002}). 
The time for a particle to go through a
cluster ranges from 7 to 10 $T.S.U.$ at the lowest available energies
($E_{c.m.}/N \leq 30 \> E.S.U/.$) and 
ranges from 3 to 5 $T.S.U.$ at the highest available energies
($E_{c.m.}/N \geq 90 \> E.S.U/.$). As seen in
\cite{CNBD2002}, fusion-like mechanisms and particle transfer mechanisms 
are observed at low \Ecm~ values ($E_{c.m.}/N \lesssim 60 \> E.S.U.$), 
while multi-fragment production and neck formation and break-up 
is dominant at high \Ecm~ values ($E_{c.m.}/N > 60 \> E.S.U.$). As it will be
seen in section \ref{s:Evst}, typical
thermalisation times for such systems ranges from 15 to 20 $T.S.U.$.}

\section{Energy deposition in final clusters.}\label{s:finaux}

In this section, we will be interested in the energy deposition in clusters  at
the simulation time $t = 200 \> T.S.U.$. At this stage, the clusters are well
separated from each other in configuration space. \mod{As it will be seen later,
these clusters are thermally equilibrated. They} can be viewed
as primary clusters. For this first simple analysis, the clusters have been
identified by using the Minimum Spanning Tree method (labeled MST) which
assumes that two particles belong to the same cluster if they are in potential
interaction, i.e. if their relative distance is below the range $r_1$ of the
potential. 

\figart{!}{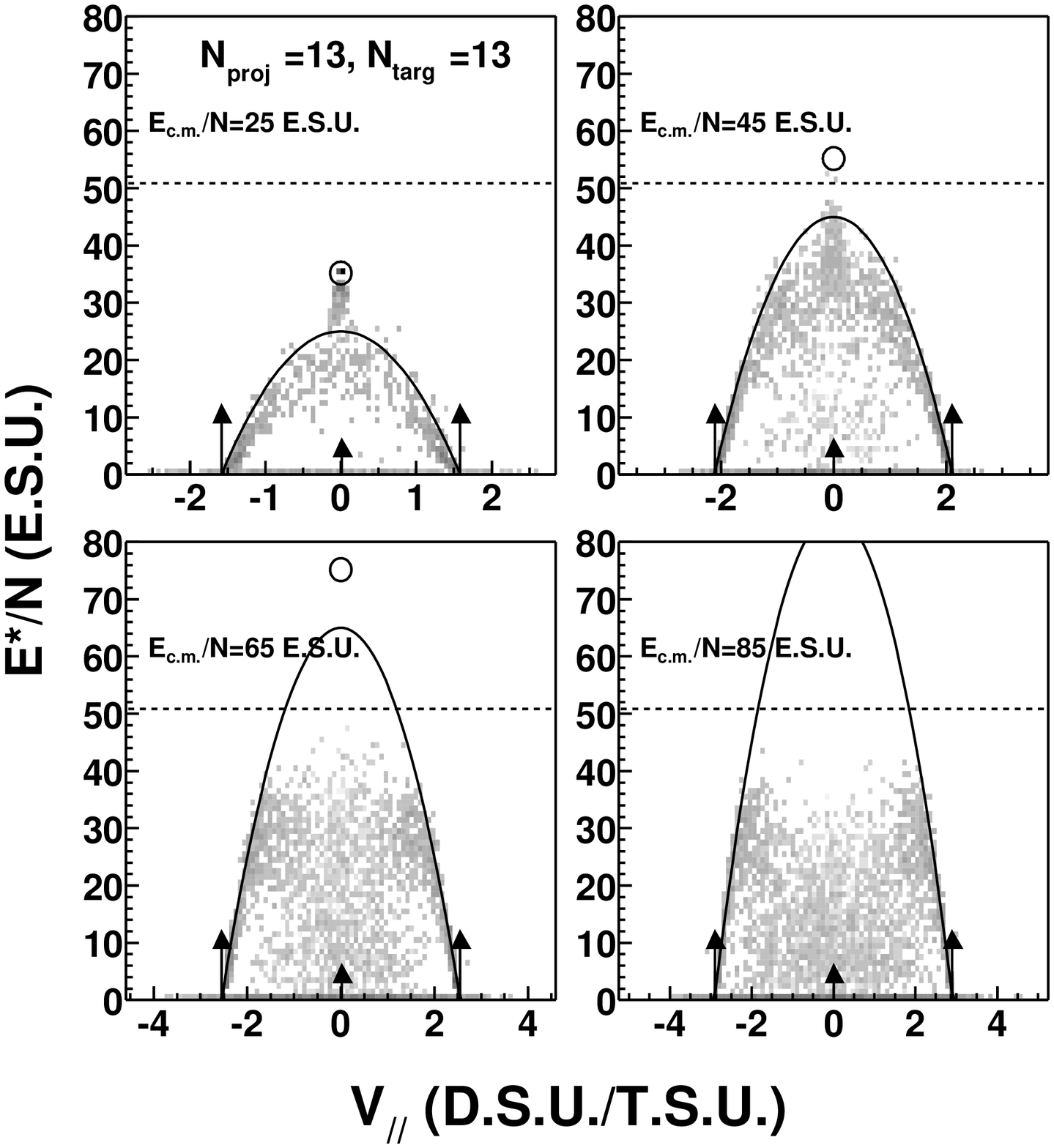}{Excitation energy of clusters versus their parallel 
velocity for \Np=13 on \Nt=13 collisions at
different available energies in the center of mass. On each panel, 
the full line corresponds to
the expected evolution for a pure binary process, 
the dashed horizontal line
corresponds to the energy of the least bound particle for $N=26$ and 
the circle corresponds
to the expected values for the fused system. 
On each plot, the darkest gray regions correspond to the highest differential
cross section values.}
{f:EvsVa13a13}{0.425}
\figart{!}{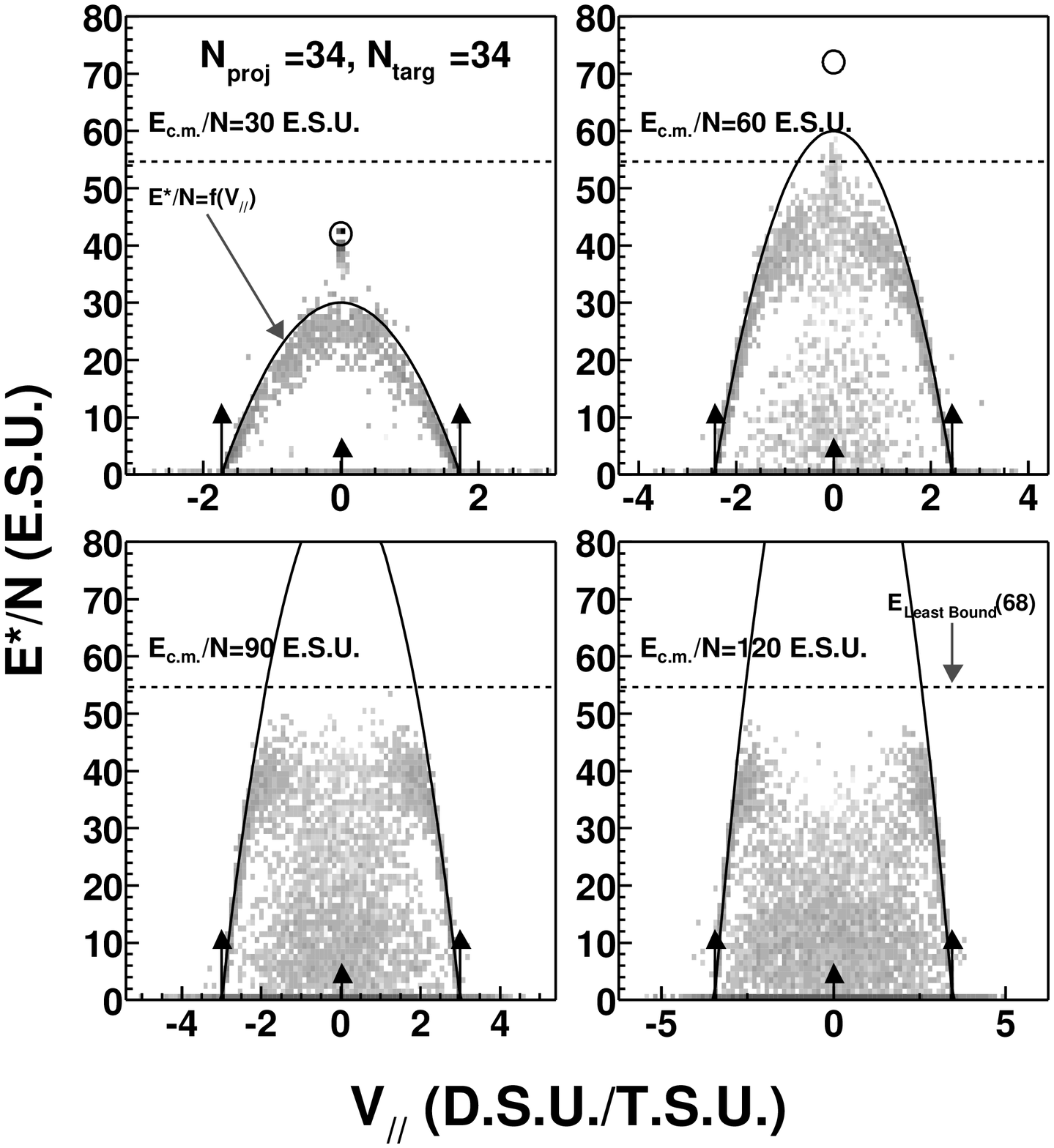}{Same as \ref{f:EvsVa13a13} 
for \Np=34 on \Nt=34 collisions 
On each panel the dashed horizontal line
corresponds to the energy of the least bound particle for $N=68$.}
{f:EvsVa34a34}{0.425}
\figart{!}{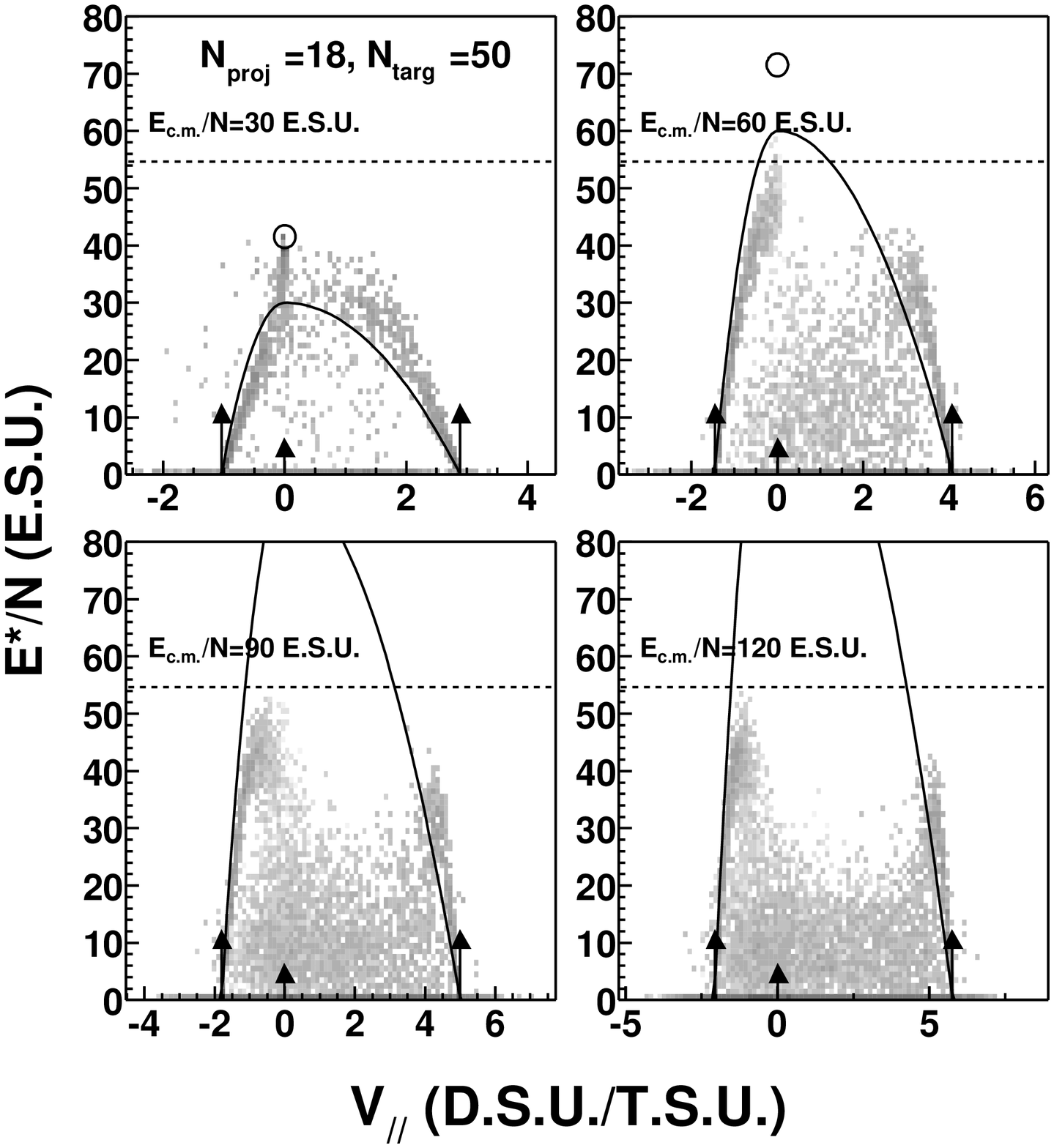}{Same as \ref{f:EvsVa13a13} 
for \Np=18 on \Nt=50 collisions 
On each panel the dashed horizontal line
corresponds to the energy of the least bound particle for $N=68$.}
{f:EvsVa18a50}{0.425}
\figart{!}{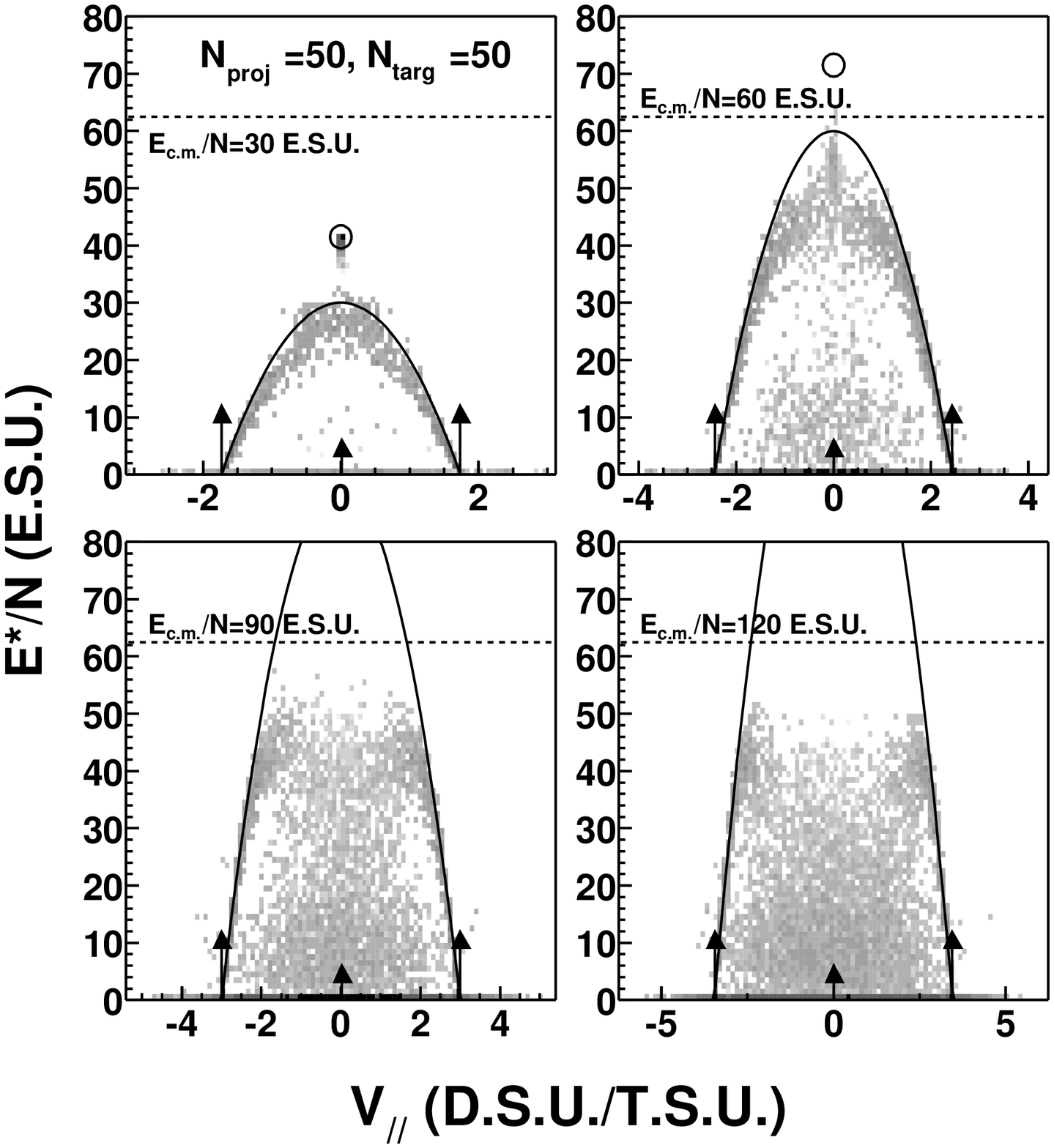}{Same as \ref{f:EvsVa13a13} 
for \Np=50 on \Nt=50 collisions 
On each panel the dashed horizontal line
corresponds to the energy of the least bound particle for $N=100$.}
{f:EvsVa50a50}{0.425}
\figart{!}{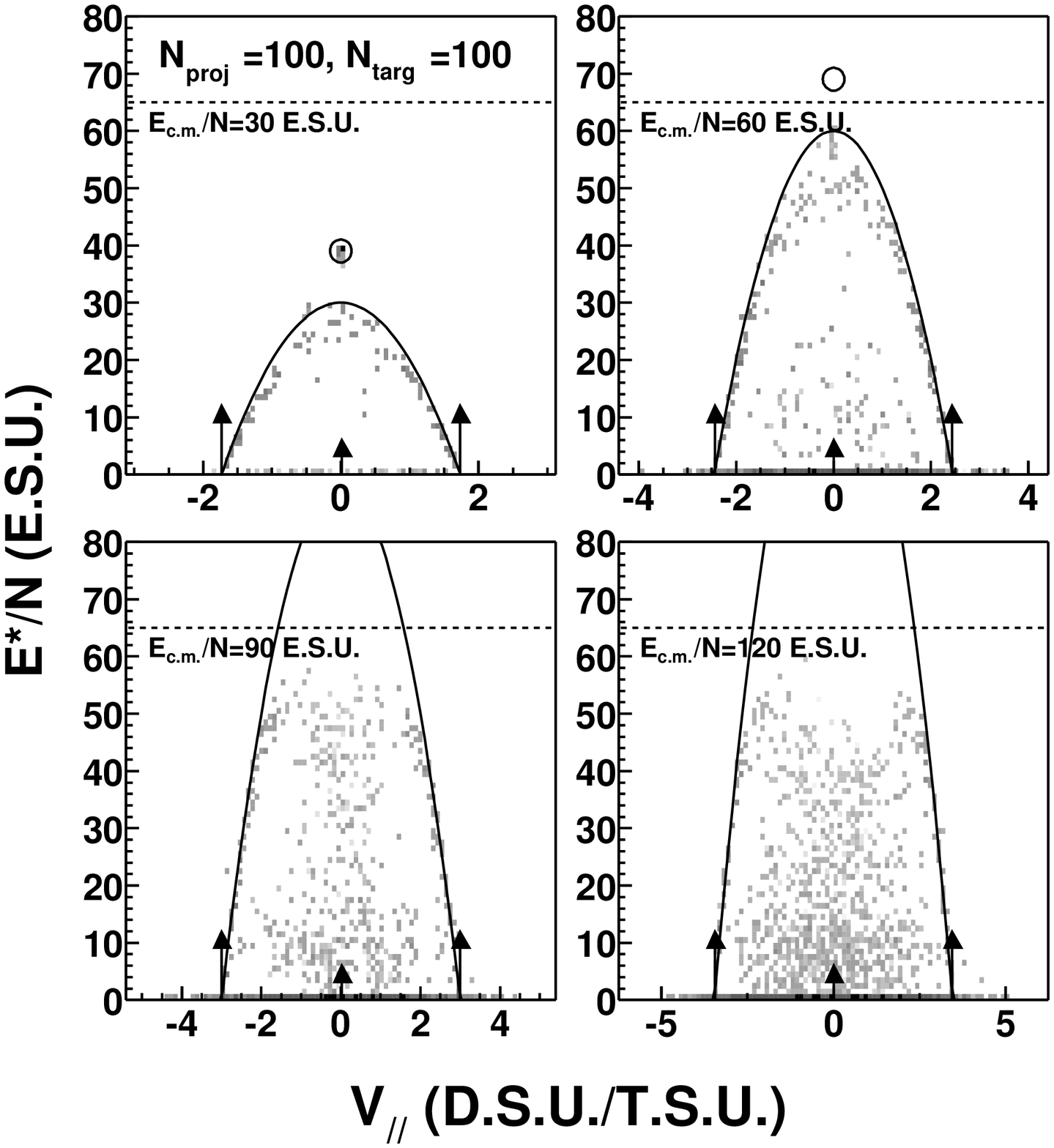}{Same as \ref{f:EvsVa13a13} 
for \Np=100 on \Nt=100 collisions 
On each panel the dashed horizontal line
corresponds to the estimated energy of the least bound particle for $N=200$.}
{f:EvsVa100a100}{0.425}

\subsection{Variations of the energy deposition with the velocity of the cluster}

On figures \ref{f:EvsVa13a13}, 
\ref{f:EvsVa34a34}, \ref{f:EvsVa18a50}, \ref{f:EvsVa50a50},
and \ref{f:EvsVa100a100} are plotted the evolution of 
the excitation energy \Estar~ 
of the cluster with its parallel velocity
\Vpar~for the whole impact parameter range and for different available
energies in the center of mass. 
The excitation energy of each cluster is
simply the difference between the total energy (potential plus kinetic) and the
``ground state'' energy of the cluster:

\begin{equation}
E^* = \sum_{i}{E_i^{kin}} + \sum_{i,j,i>j}
{V(|\overrightarrow{r_{i}}-\overrightarrow{r_{j}}|)} - E_{bind}(N)
\end{equation}

\noindent where $E_i^{kin}$ is the kinetic energy of the particle $i$ in the
cluster's centre of mass, $V(|\overrightarrow{r_{i}}-\overrightarrow{r_{j}}|)$
the potential energy between the particles $i$ and $j$, $E_{bind}(N)$ the
binding energy of the cluster and $N$ its number of particles. This excitation
energy is determined at the end of the calculation corresponding to  $t=200 \>
T.S.U.$. As it will be seen in the next section, this energy is very  close to
the one obtained at the  separation time of the clusters (the smallest time at
which clusters can be identified),  since in this time range the emission of
monomers or small clusters by the primary clusters (evaporation) is
very weak and the clusters have no time to cool down significantly
\cite{Dorso99}.  On each panel of the figures, the full line corresponds to the
expected correlation between \Estar~and \Vpar~for a pure binary scenario, i.e.
the excitation energy is only due to the velocity damping of each partner,
without any particle exchange between them. This excitation energy is then
determined by:

\begin{equation}
E^*=\frac{1}{2} \frac{m_{p} m_{t}}{m_{p}+m_{t}} 
\left(v_{rel,max}^2-v_{rel}^2\right)
\end{equation}

\noindent where $m_{p}$ and $m_{t}$ are the projectile and the target mass
respectively, $v_{rel,max}$ the initial relative velocity between the
projectile and the target and $v_{rel}$ is the relative velocity between the
projectile and the target after interaction. The mass of each cluster is
proportional to the number of particles $m(N)=m_{particle}\times N$ where
$m_{particle}=20\> S.U.$. The initial relative velocity is given by:

\begin{equation}
v_{rel,max}=\sqrt{2 E_{c.m.} \frac{m_p+m_t}{m_p m_t}}
\label{e:vrelmax}
\end{equation}

\mod{The
small circle displayed on each panel is centered around the 
expected values of velocity and excitation energy for the fused system.
The horizontal dashed line corresponds to  the energy  of the
least bound particle $E_{LeastBound}$ 
for the fused system $N=N_{proj}+N_{target}$.  
This energy is determined for the stable clusters (``ground state'') 
and is defined as follows:
\begin{equation}
E_{LeastBound}=\max_{j=1,N} \left[\sum_{i\neq j} V(r_{ij}) \right]
\end{equation}
\noindent where $r_{ij}$ is the relative distance between the particles $i$ and
$j$ and $V(r_{ij})$ is the value of the two-body potential. As it has been shown
in \cite{CNBD2002}, this energy varies dramatically with $N$. 
In their ``ground states'' the clusters are small crystals and the \Elb~ value 
is mainly due to geometrical effects (number of neighbors of a particle at the
surface). Such variations are well known in cluster physics \cite{WALES97}. 
}

At low \Ecm~values ($E_{c.m.}/N=25\> E.S.U.$ for the 13+13 system and
$E_{c.m.}/N=30 \> E.S.U.$ for the others), the points are slightly below the
full line. This means that the excitation energy is strongly linked to the
velocity damping. In that case, the collisions leads to the formation of 
excited projectile-like and target-like clusters. 
The small shift is due to mass transfers between the
projectile and the target, and to promptly emitted clusters.  The area
corresponding to the fused system is well populated showing that a complete
fusion process occurs. At intermediate energies ($E_{c.m.}/N=45\> E.S.U.$ for
the 13+13 system and $E_{c.m.}/N=60 \> E.S.U.$ for the others), the
distribution of points is roughly compatible  with the pure binary process
hypothesis (formation of excited projectile-like and target-like clusters) 
except  for cluster velocities which lead to excitation energies per
particle higher than $E_{LeastBound}(N_{fused})$ (where
$N_{fused}=N_{proj}+N_{target}$) in the pure binary process picture. For
these clusters, $E^*/N$ is always smaller than  $E_{LeastBound}(N_{fused})$. 
The complete
fusion process  area, located above $E_{LeastBound}(N_{fused})$, 
is empty. In that case, as it will be seen in the next section, incomplete
fusion process occurs. For the two
highest energies, this trend is enhanced. Around the projectile and the target
velocity the projectile-like and target-like 
clusters have an excitation energy compatible with the pure binary
process hypothesis.  Around the centre of mass velocity, when this picture 
would give excitation energies per particle higher than 
$E_{LeastBound}(N_{fused})$, one
finds clusters at small excitation energies. The energy of the least bound 
particle seems to be a limit to the excitation energy which can be stored in
these classical clusters.

\figart{!}{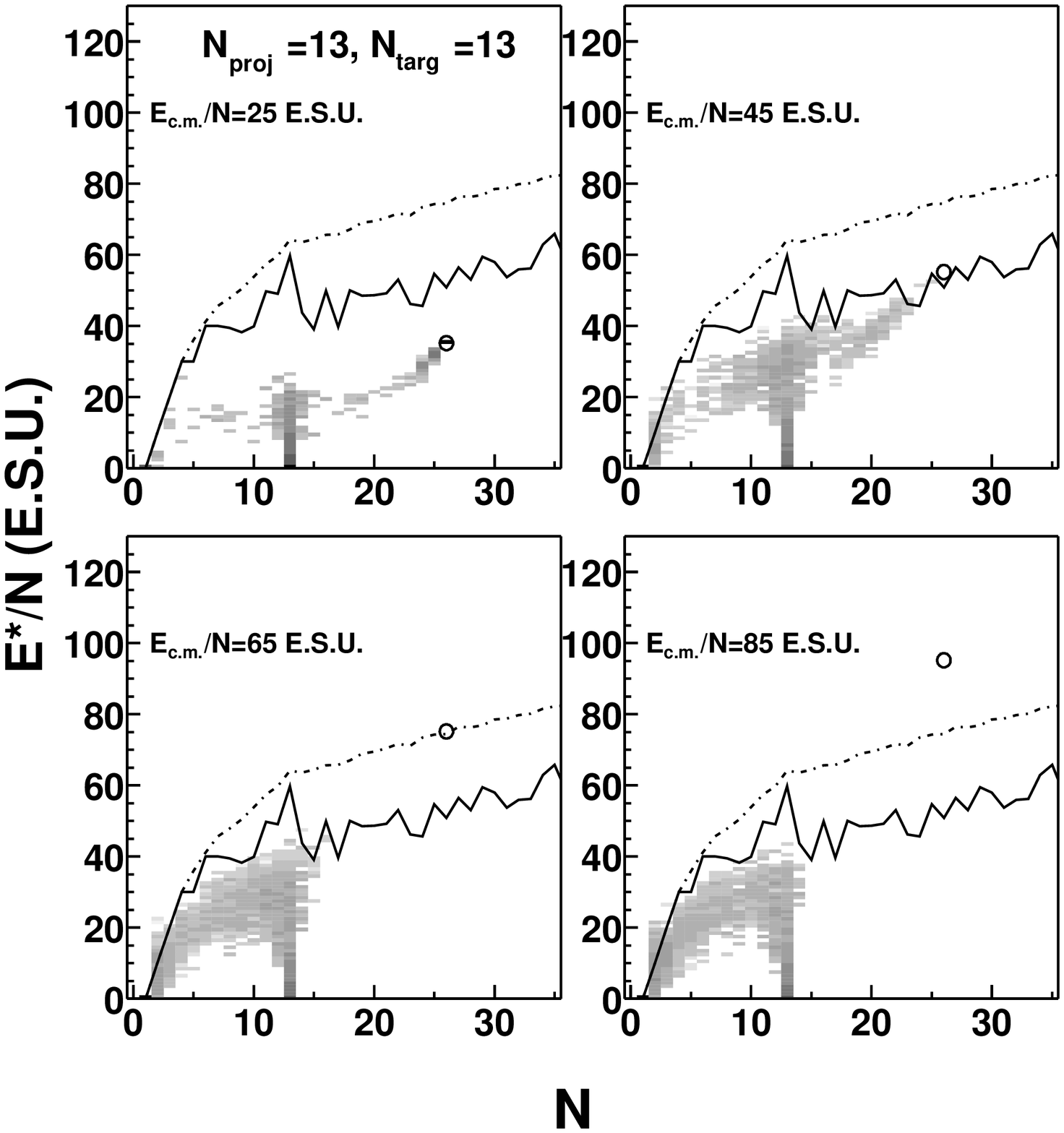}{Excitation energy of clusters versus $N$ 
for \Np=13 on \Nt=13 collisions at
different available energies in the center of mass. On each panel, 
the full line corresponds to
the energy of the least bound particle, 
the dashed and dotted line
corresponds to the binding energy per particle and 
the circle corresponds
to the expected values for the fused system. 
On each plot, the darkest gray regions correspond to the highest differential
cross section values.}
{f:EvsNa13a13}{0.425}
\figart{!}{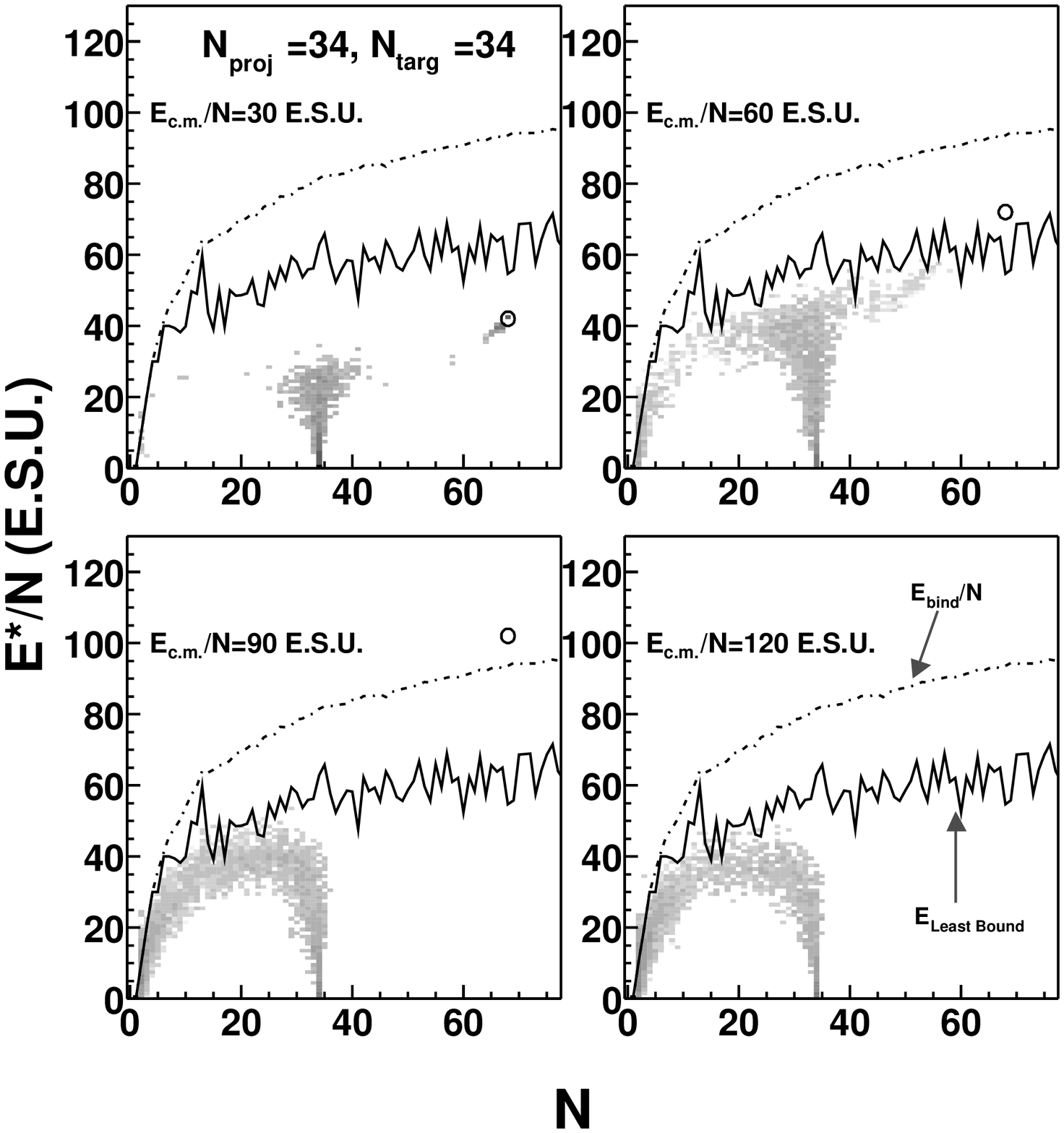}{Same as \ref{f:EvsNa13a13} 
for \Np=34 on \Nt=34 collisions.}
{f:EvsNa34a34}{0.425}
\figart{!}{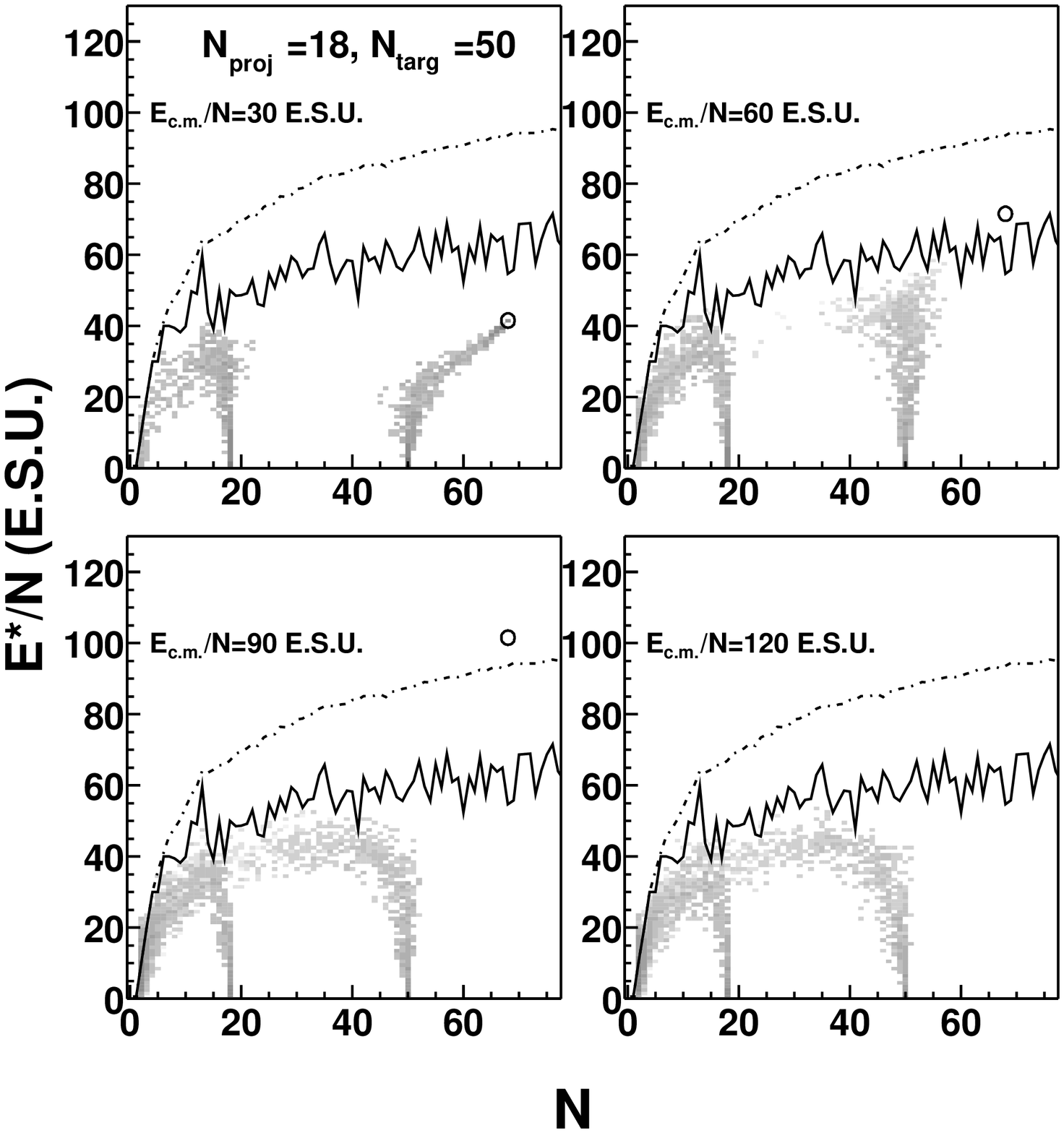}{Same as \ref{f:EvsNa13a13} 
for \Np=18 on \Nt=50 collisions.}
{f:EvsNa18a50}{0.425}
\figart{t}{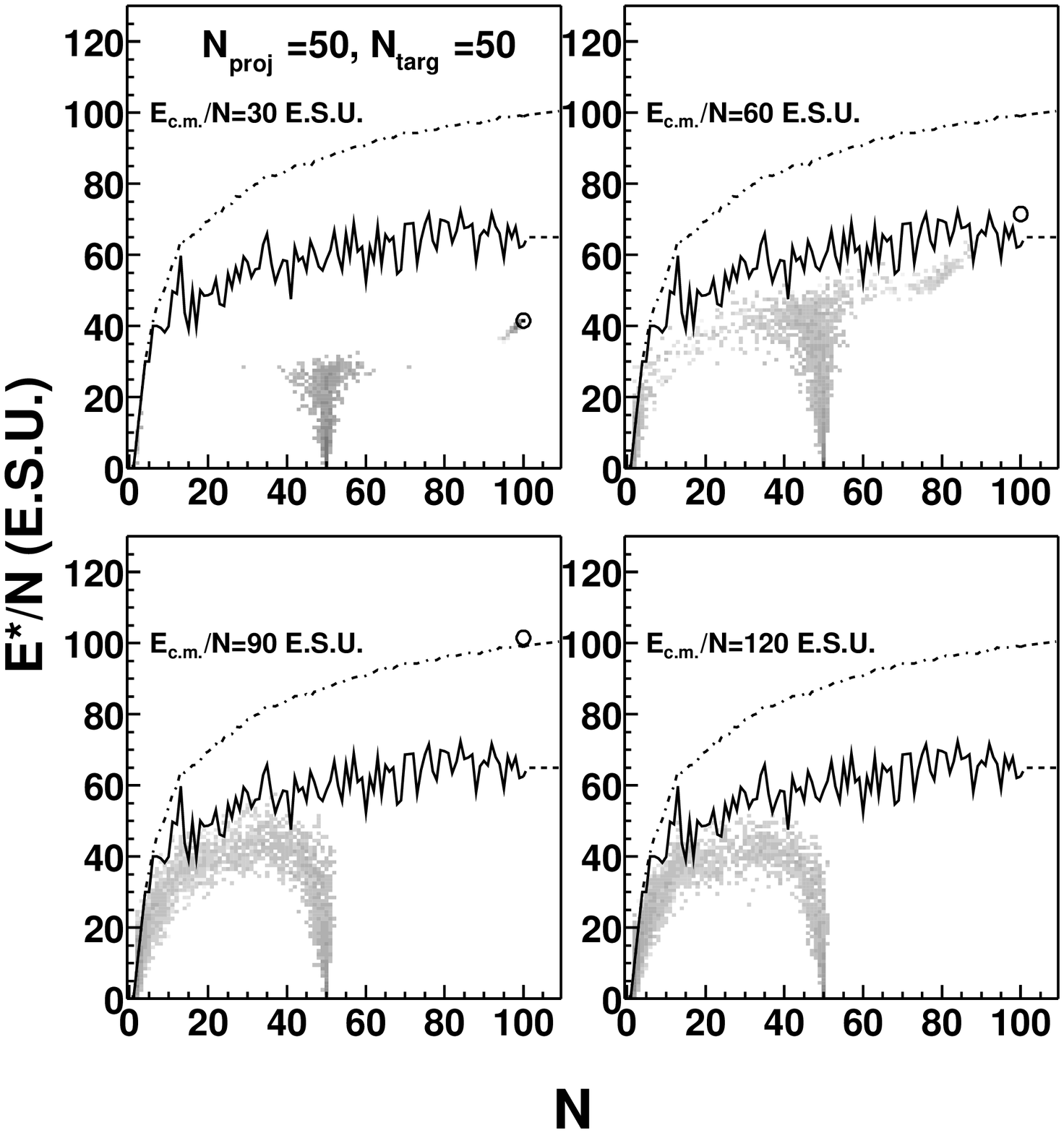}{Same as \ref{f:EvsNa13a13} 
for \Np=50 on \Nt=50 collisions.}
{f:EvsNa50a50}{0.425}
\figart{t}{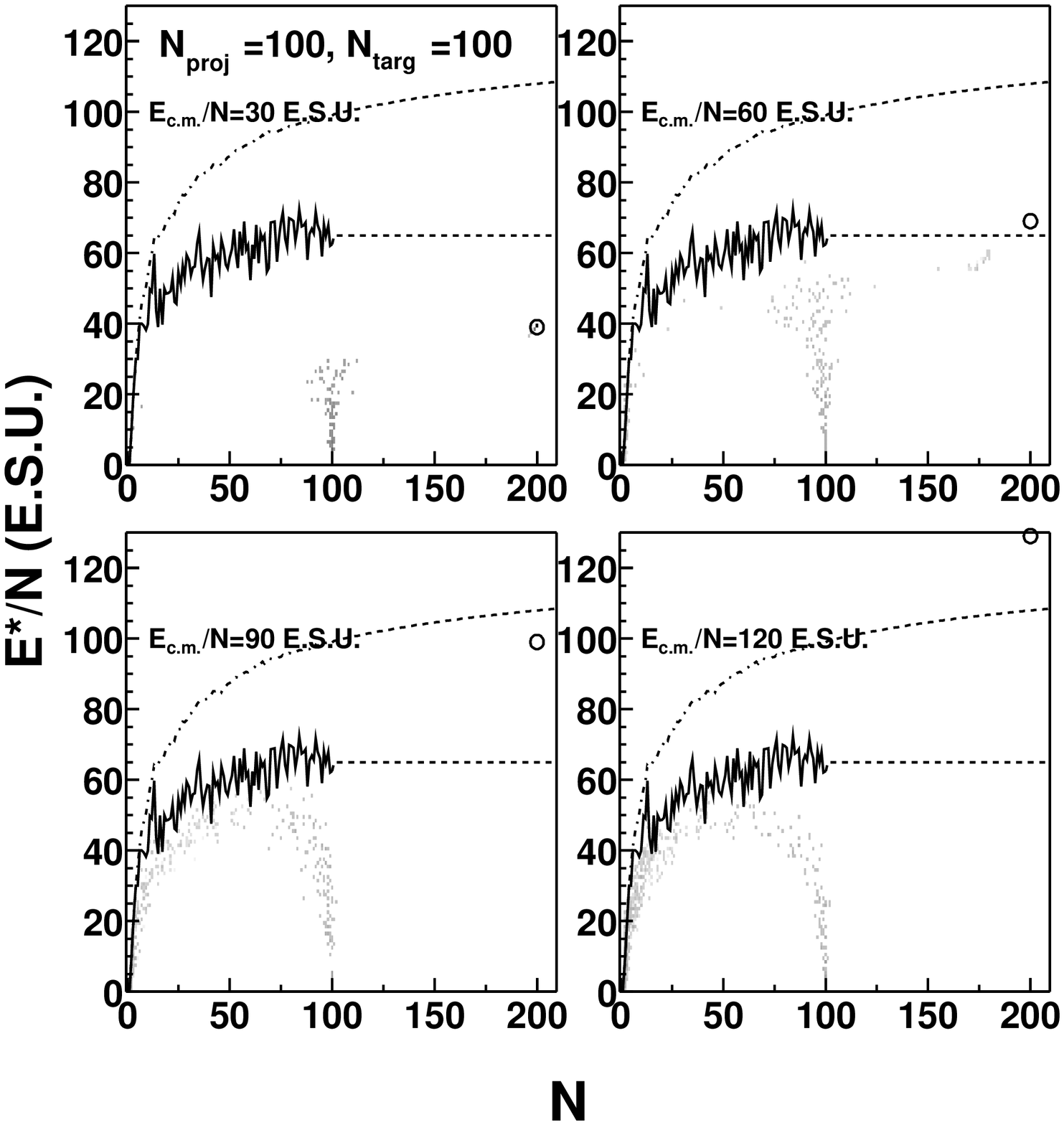}{Same as \ref{f:EvsNa13a13} 
for \Np=100 on \Nt=100 collisions. The dashed lines on each panel correspond to
the extrapolated values of the binding energy (upper line) and the extrapolated
energy of the least bound particle (lower line) for $N > 100$.}
{f:EvsNa100a100}{0.425}

\subsection{Variations of the energy deposition with the size of the cluster}

This limitation can be more clearly seen when the excitation energy \Estar~is
plotted  as a function of $N$,  as in figures \ref{f:EvsNa13a13}, 
\ref{f:EvsNa34a34}, \ref{f:EvsNa18a50}, \ref{f:EvsNa50a50} and
\ref{f:EvsNa100a100}. On each panel, the full line corresponds to the energy of
the least bound particle \Elb~in the cluster and the dashed line to the binding
energy per particle \Ebind. As in figures \ref{f:EvsVa13a13}, 
\ref{f:EvsVa34a34}, \ref{f:EvsVa18a50}, \ref{f:EvsVa50a50}, and
\ref{f:EvsVa100a100}, the small circle corresponds to the expected values for
the fused system. At low energy, the area corresponding to complete fusion is
filled and all the available energy can be stored as excitation energy. But for
higher energies, one can clearly see that for each fragment size, \Estar~never
overcomes \Elb. At intermediate energy, clusters with sizes higher than the
projectile size and the target size can be seen. This area corresponds to an
incomplete fusion process. For the two highest energies, the plots are almost
identical: there is no more fusion (the small circle area is empty) 
and the clusters are smaller than the target
and than the projectile.  One can notice that \Estar~never reaches the binding
energy \Ebind~except for small clusters where \Ebind~and \Elb~are equal. The
energy of the least bound particle in the cluster  \Elb~is an upper limit for
the energy per particle which can be stored in the cluster, whatever the system
size, the available energy in the center of mass and the asymmetry of the
entrance channel.

\figartlarg{htb}{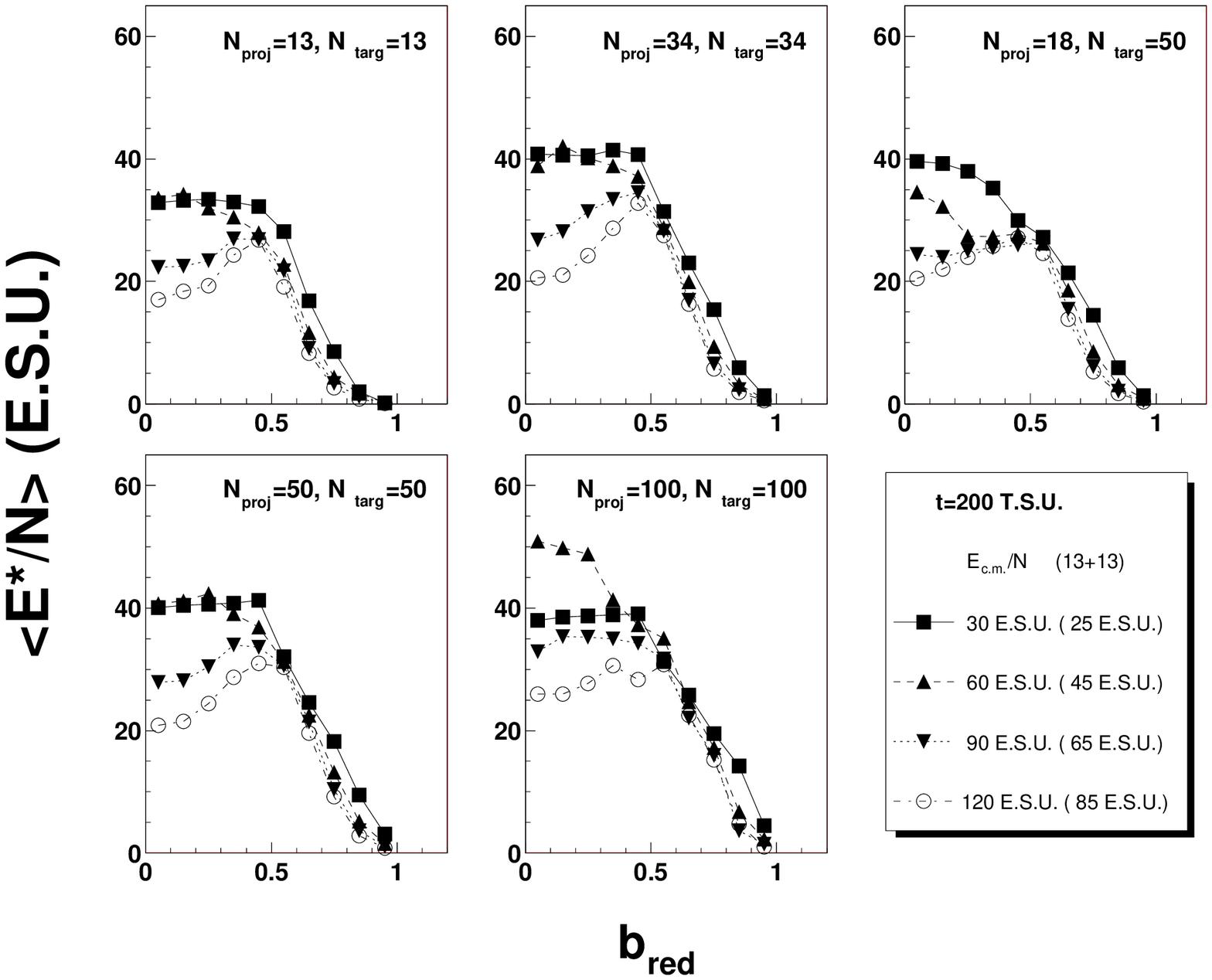}{Average excitation energy stored in the clusters
as a function of the reduced impact parameter. The different lines and symbols
correspond to different values of the available energy per particle in the
center of mass frame. The \Ecm values indicated between parentheses correspond
to the 13 + 13 system. See text for details.}{f:tEvsb}{0.85}

\figartlarg{htb}{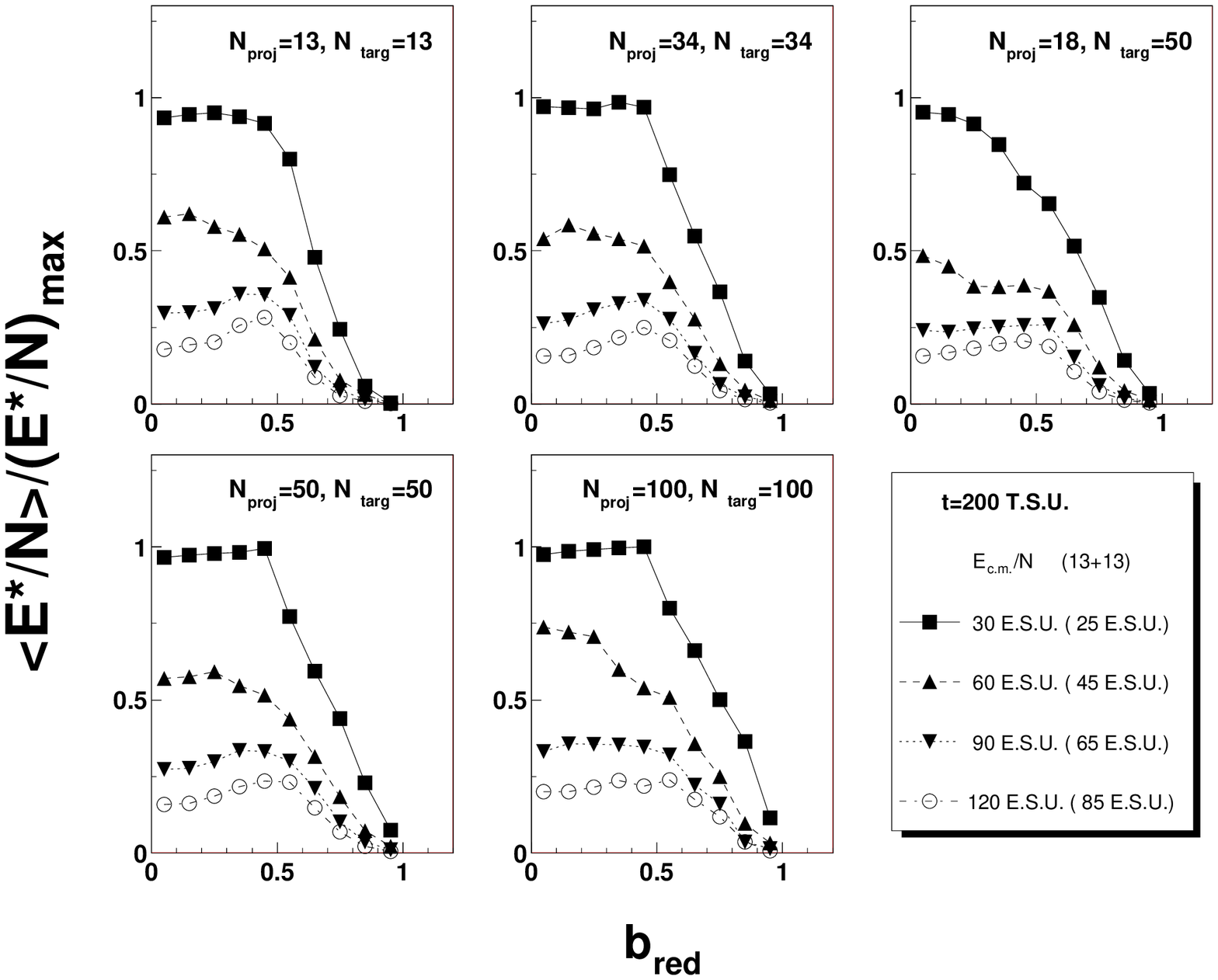}{Average amount of available energy stored in the
clusters as a function of the reduced impact parameter. The different lines and symbols
correspond to different values of the available energy per particle in the
center of mass frame. The \Ecm values indicated between parentheses correspond
to the 13 + 13 system. See text for details.}{f:tErelvsb}{0.85}

\subsection{Variations of the energy deposition with the impact parameter 
and the available energy in the center of mass.}

Figure \ref{f:tEvsb} shows the variations of the average 
excitation energy per particle $<E*/N>$ with the reduced impact parameter
$b_{red}=b/b_{max}$ for different
available energies in the center of mass frame \Ecm. The average is calculated
for clusters with $N$ greater or equal to 3. Each panel corresponds
to a system. The squares 
correspond
to the lowest \Ecm~values, the triangles to \Ecm~values close to
\Elb, the diamond to \Ecm~values close to
\Ebind~and the circles to the highest \Ecm~values. As in the
previous figures, the average excitation energy is limited around $40\>E.S.U.$
except for the 100 + 100 system for which the maximum energy reached is $50\>
E.S.U.$ at $E_{c.m.}/N=60\> E.S.U.$. 

At the lowest energy, for symmetric systems,  
$<E*/N>$ increases when \bred~decreases down to
$b_{red}=0.4$, and then is constant at $<E*/N>\approx40 E.S.U.$ below.  This
saturation is due to the occurrence of fusion: the maximum excitation energy
is reached by the fused system. For the asymmetric system 18 + 50, saturation
occurs at smaller impact parameters. For intermediate energies, the picture is
roughly the same. But although the available energy is higher, the maximum
$<E*/N>$ value is almost the same as for the lowest available energy. 
In that case, an
incomplete fusion process occurs and the excitation energy of the heaviest
cluster is limited by the energy of its least bound particle, which is well
below the expected excitation energy of the fused system. For the two highest
energies, $<E*/N>$ increases when \bred~decreases down to $b_{red}=0.4$ and
then $<E*/N>$ decreases when \bred~decreases. This effect results from the
lower cluster sizes for the most central collisions, which at these energies
produce several clusters of intermediate sizes. Indeed, the smaller is the
cluster size, the lower is the upper limit in energy storage. Since clusters
have a smaller size for central collisions, the $<E*/N>$ decreases
consequently. This evolution does not seem to depend strongly on the
total system size or on the entrance channel asymmetry.

The same evolution is seen on figure \ref{f:tErelvsb} which shows the evolution
of the ratio $<E*/N> / (E*/N)_{max}$ where $(E*/N)_{max}$ corresponds to the
maximum excitation energy expected for the fused systems. For low energies,
this ratio continuously increases when \bred~decreases and is close to 1 below
$b_{red} \approx 0.4$: this corresponds to the occurrence of the fusion process.
This value of 1 is reached for \bred~below 0.2 for the 18 + 50 system. This
difference can be understood quite easily: for a fixed \Ecm~value, the relative
velocity between the projectile and the target is higher for an asymmetric
system than for a symmetric one (see equation \ref{e:vrelmax}).  This leads to
an interaction time smaller for the asymmetric system than for a symmetric one
and then to less efficient energy deposition in the clusters. For higher
energies, this value of 1 is never reached. The maximum value even decreases
when \Ecm~increases: the relative amount of energy stored in clusters is lower
and lower when the available energy increases.

\subsection{Discussion}

These studies show that in classical systems, the energy storage in clusters
is limited. Such a saturation was observed for two-dimensional classical systems
\cite{Strachan98}.
This limitation is only linked to the intrinsic properties of the
cluster: the energy deposition cannot be higher than the energy of the least
bound particle in the cluster.

This limitation of excitation energy can be understood quite easily.  The least
bound particle remains bound to the cluster only if its total energy is
negative, i.e. its kinetic energy due to the excitation  is below its potential
one. If one assumes that the excitation energy is roughly equally shared over
all particles in the cluster, when the kinetic  energy balances the potential
energy of the least bound particle,  this particle is no \mod{longer}
bound to the
cluster and quickly escapes. To be observed for a long time (greater than the
thermalisation time), the excited cluster
must have an excitation energy  per particle below the energy of the least
bound particle.

The mechanism of energy deposition in classical N-body clusters seems to be the
following one: the excitation is mainly driven by the velocity damping of the
two partners and to a lesser extent by exchanges of  particles between them.
Once the energy of the least bound particle is reached, unbound particles
and/or clusters escape quickly and keep an excitation energy per particle below
the energy of the least bound one. As a consequence, the highest energy
deposition per particle  can only be obtained at \Ecm~energies close to \Elb. 
For higher available energies, the system fragments quickly, leaving rather
``cold'' clusters around the center of mass velocity. The energy in excess is
stored in the kinetic energies of the clusters. A similar mechanism could be 
an explanation of the apparent saturation of primary fragments excitation
energies observed for central collisions of
the Xe + Sn system when the incident energy increases from 32 to 50 A.MeV 
\cite{Marie98,Hudan}. This could also be an
explanation for the limitation of temperature values around 5 MeV  by using
isotope ratios methods or population ratios methods \cite{Ma98,Poch95}. Such a
limitation has been observed  in the fragmentation of Uranium projectiles at
relativistic energies
\cite{Schmidt2002}. This limitation was also suggested by the
observation of the saturation of the evaporated neutron multiplicity when the
incident  energy increases \cite{Galin94}. 

Providing this conclusion can be applied in nuclear physics, i.e. \mod{if}
quantum effects and Coulomb interaction do not modify strongly the above 
picture, this
means that the maximum energy per nucleon which can be stored in
\mod{thermalised} hot nuclear fragments 
would never exceed the energy of the least bound nucleon.  This energy would
correspond to the energy of the last populated level. If this assertion is
true, that would be in contradiction with several experimental works in which
high energy depositions (up to 20 A.MeV) were measured
\cite{Borderie96,Ma98,Nato2002}. \mod{This discrepancy may result from the
calorimetry analyses used in these articles, 
whereas $E*/N$ is directly extracted in this work.}

One has to be careful because the observations made here are for a
fixed simulation time. At the early stages of the collision higher energy
deposition could be reached for short times. 
The next section will be devoted to
the evolution of this energy deposition with the reaction time.

\figartlarg{htb}{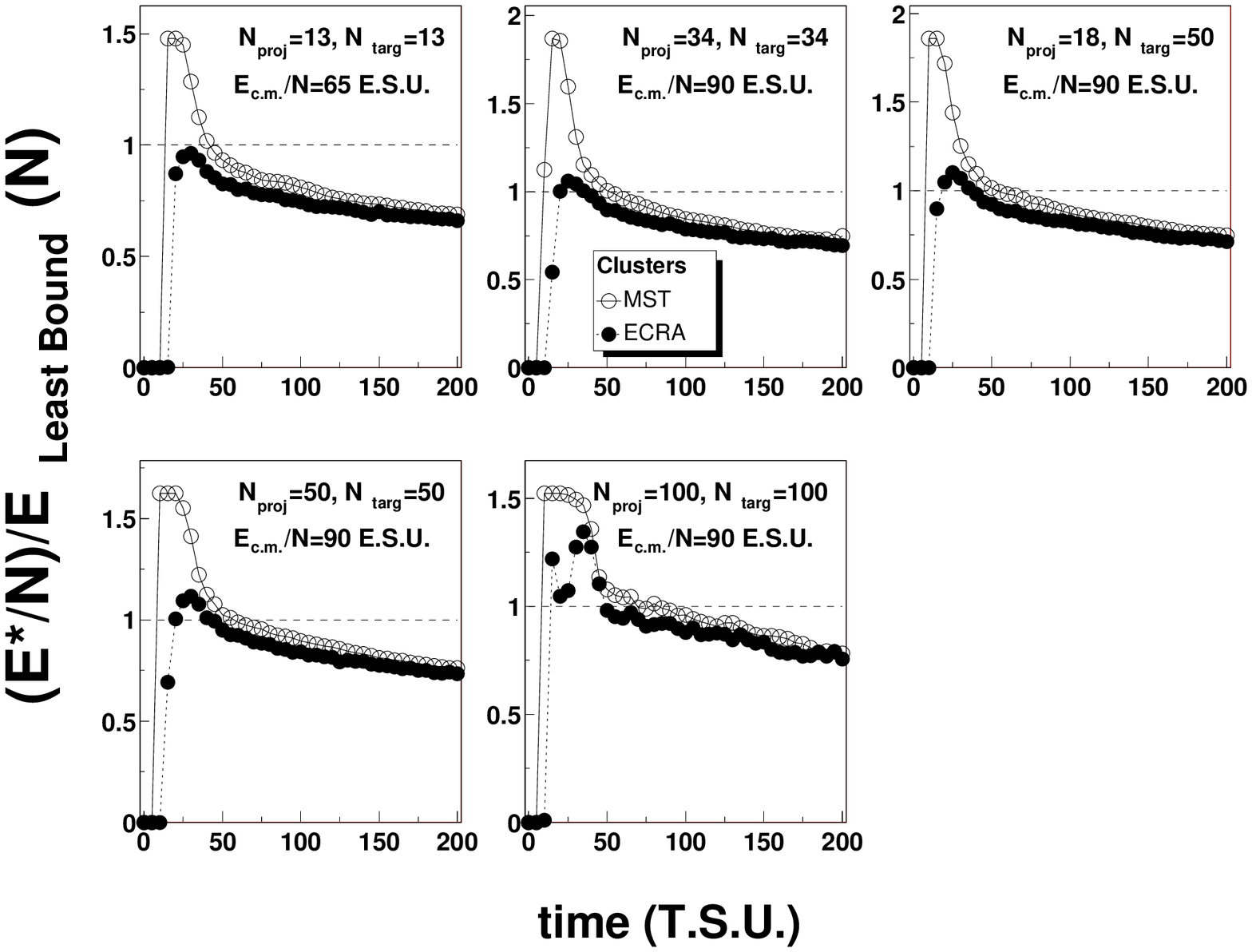}{Variation of the ratio $(E^*/N)/E_{LeastBound}$ with
time for various systems and for central collisions ($b_{red} < 0.1$) 
when the available energy in the centre of mass is
close to the binding energy of the fused system. The open circles correspond to
the MST clusters and the full circles to the ECRA clusters.}{f:Erelvst90}{0.85}

\figartlarg{htb}{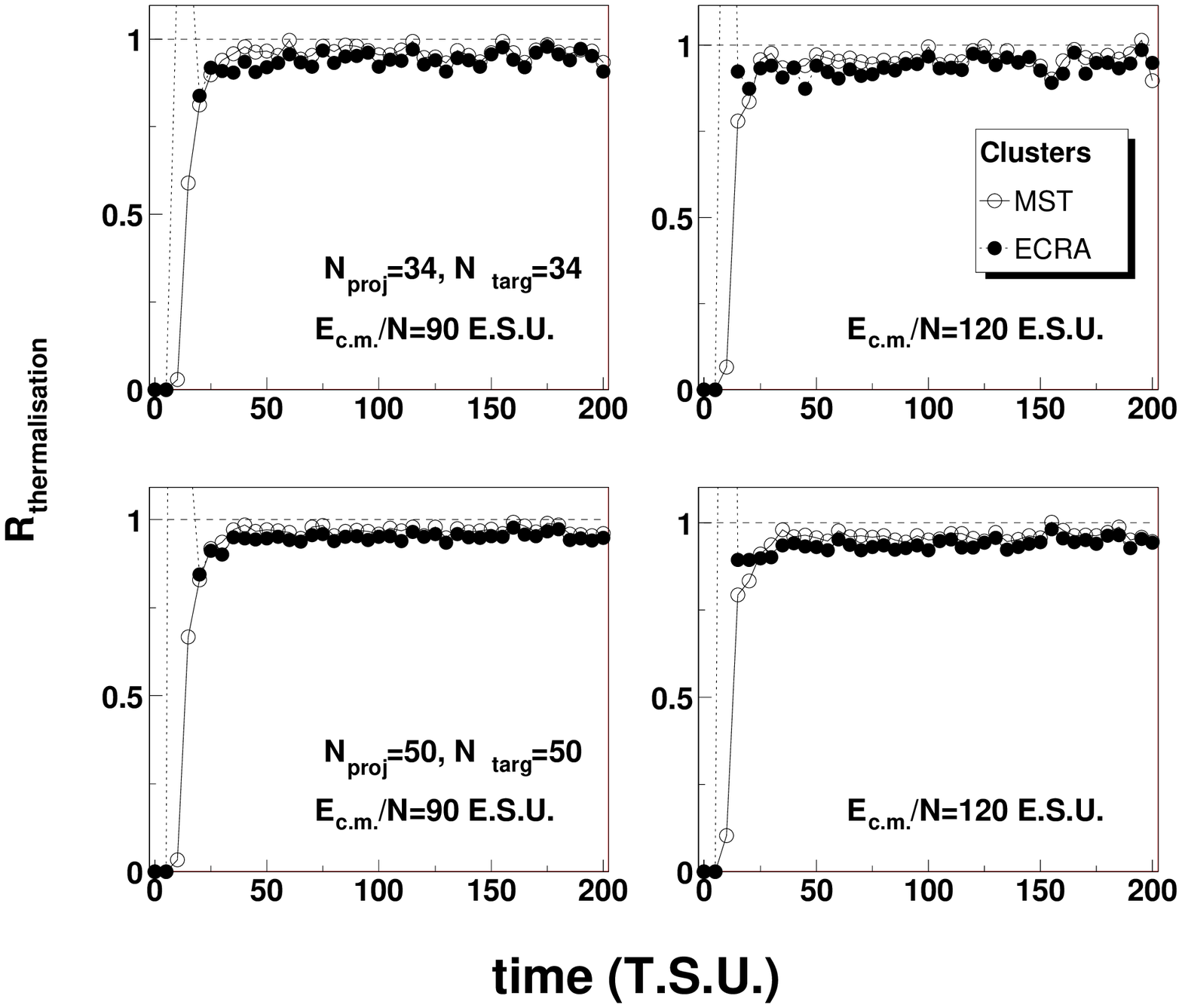}{Variation of the ratio 
$(\sigma(E_{kin})/<E_{kin}>)/\sqrt{\frac{2 N}{3 N-2}}$ with time for the 34+34
system (first raw) and the 50+50 system (second raw), for 
central collisions ($b_{red} < 0.1$) at the two highest \Ecm~
values (90 $E.S.U.$ left column and 120 $E.S.U.$
right column). The open circles correspond to
the MST clusters and the full circles to the ECRA clusters.}{f:thermal}{0.85}

\figartlarg{htb}{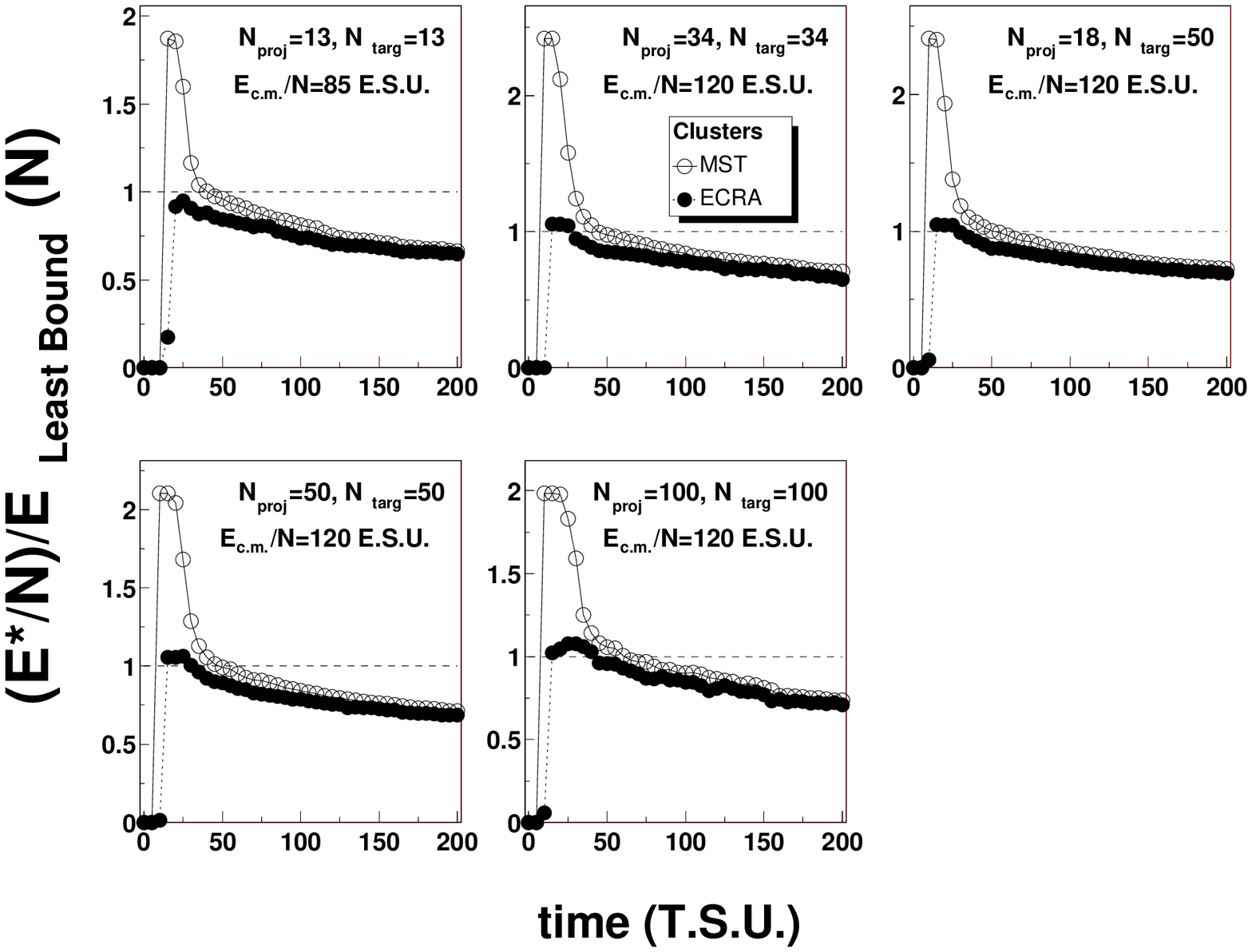}{Variation of the ratio $(E^*/N)/E_{LeastBound}$ with
time for various systems and for central collisions ($b_{red} < 0.1$) 
when the available energy in the centre of mass is
close to the most bound particle in the fused system. The open circles correspond to
the MST clusters and the full circles to the ECRA clusters.}{f:Erelvst120}{0.85}

\section{Evolution of the energy deposition with time.}\label{s:Evst}

The aim of the present section is to check to which extent the limitation  of
the energy deposition observed for final (long simulation times) 
\mod{thermally equilibrated} clusters is
true \mod{during} the collision. 
Since the limitation is linked to the energy of
the least bound particle in the cluster, we will follow the ratio of the
excitation energy per particle to the energy of the least bound particle
$(E*/N)/E_{LeastBound}(N)$ with collision time.  

Such an evolution is shown in figure \ref{f:Erelvst90} for MST clusters  for
central collisions \mod{($b_{red} < 0.1$)} for all systems at 
$E_{c.m.}/N=90  \> E.S.U.$  ($65  \> E.S.U.$ for the 13 + 13 system). It 
is seen that for MST clusters (open
circles) very high energies can be found, up to the maximum excitation energy.
\mod{This means that very excited systems could be formed at the early stages
of the collision. But one has to check first if this transfered energy is
equally shared between all degrees of freedom and hence if these intermediate
clusters are thermalised.}

\mod{One way to check this thermalisation is to follow the ratio  
between the dispersion of the distribution of the kinetic energy of 
particles
$\sigma(E_{kin})$ in the cluster and its average value
$<E_{kin}>$. For a thermalised system, the value of this ratio is well
defined: it is equal to $\sqrt{2/3}$ in the canonical ensemble, and equal to
$\sqrt{\frac{2N-2}{3N-1}}$ in the micro-canonical ensemble \cite{Lopez89}, 
where $N$ is the
number of particles in the cluster. The time evolution of  
$R_{thermalisation}=(\sigma(E_{kin})/<E_{kin}>)/\sqrt{\frac{2N-2}{3N-1}}$ 
is displayed on figure 
\ref{f:thermal} for the 34+34 system (first row) and the 50+50 system (second
row) at the two highest \Ecm~ values (90 $E.S.U.$ left column and 120 $E.S.U.$
right column). A value close to 1 indicates that the cluster is thermalised. 
Let us first follow the evolution of the ratio for MST clusters
(open circles). The thermalisation of clusters is reached at $t\approx 50 \>
T.S.U.$ for both systems at \Ecm=90 $E.S.U.$. 
The excitation energy per particle of the MST
clusters at this time (see figure \ref{f:Erelvst90}, open circles) is close to
or below the energy \Elb~ of the least bound particle in the cluster. 
 For later times, the ratio
slightly varies around 1, showing that clusters stay thermalised even if
evaporation of light clusters occurs.
The same
conclusion can be drawn at \Ecm=120 $E.S.U.$ for which the thermalisation is
reached at $t\approx 35\>T.S.U.$ when the excitation energy per particle 
is very close to \Elb. From this simple study, one can conclude that the highly
excited intermediate cluster is not thermalised, and that thermalised clusters
have an excitation energy per particle close to or below \Elb.}

\mod{It} has been found in earlier works that such systems are indeed composed 
of several small clusters \mod{at early times} 
\cite{Aichelin,Dorso97,Sator2001}. 
The size distributions of
these clusters have been  found to be almost identical to those identified after
their separation \cite{Sator2001}. The problem is how to identify these clusters.
In that respect, the MST algorithm is no longer suited since particles close to
each other at these early stages are not necessarily bound at later times.
Different methods have been developed to identify clusters. One is 
\mod{to assume}
that two particles are \mod{belong} to the same cluster if they are bound, i.e. if
their relative energy is negative. These clusters have been labeled
Coniglio-Klein clusters (CK clusters) or MSTE clusters in other works. Another
way to define clusters is to find at each time step the most bound partition of
particles in clusters. This algorithm is described in detail in \cite{ECRA} and
is \mod{called} 
Early Cluster Recognition Algorithm (ECRA). The latter will be used in
the present work.

On figure \ref{f:Erelvst90} the evolution of $(E*/N)/E_{LeastBound}$ with the
collision time is shown for ECRA (\mod{filled} circles) clusters. 
Above $t\approx 
50  \> T.S.U.$ both algorithms (MST and ECRA) give the same results: 
clusters are well
separated in the configuration space. At this time, clusters do not interact
anymore with each other: this corresponds to the ``freeze-out'' configuration
assumed in statistical multifragmentation models \cite{SMM,MMMC}. \mod{One has
to notice that this time coincides with the thermalisation time found in the
previous paragraph. The thermalisation of ECRA clusters
(see figure \ref{f:thermal}, filled circles) is
also reached at the ``freeze-out'' time}. After this
time, clusters have an excitation energy which is below \Elb~(ratio below 1).
But for earlier times, the configuration is more complex. While MST clusters
reach very high excitation energies per particle, the ratio for  ECRA clusters
is always around or below 1, except for the 100+100 system. Apart for the
latter system, the limitation observed in the previous section seems to be still
true all along the collision time. But what leads to the difference observed for
the 100+100 system? 

Indeed, for such a heavy system, pre-formed clusters 
remain long enough together
to interact strongly. If one of the particles 
escapes from one of these clusters,
the probability that it is captured by another cluster is high. In that respect
the clusters are acting as a confining wall and inhibiting 
quick escapes of the most
energetic particles. \mod{These interactions between pre-formed clusters 
prevent their thermalisation before the ``freeze-out'' time}. 
It can be seen in figure 
\ref{f:Erelvst90} that the maximum value
of the ratio is higher for heavier systems. This effect 
should be reduced when the time during which clusters interact is
reduced,i.e. when the available energy increases. This is observed in figure 
\ref{f:Erelvst120} which shows the evolution of the ratio with time for
$E_{c.m.}/N=120  \> E.S.U.$ ($E_{c.m.}/N=85  \> E.S.U.$ for the 13 + 13 system). At
this energy, the reaction times are smaller and the ratio is always close to or
below 1. 

It could be argued that the observed evolution of ECRA clusters is only due 
to the
algorithm of cluster recognition used. Since one tries to find the most bound
partition, this could minimize the energies of clusters. It is indeed
observed that ECRA clusters have an excitation energy below the MST clusters.
One can make two remarks. The first one is that the \Elb~has not been used to
determine the ECRA clusters. There is hence no reason to find \Elb~as an upper
limit for the excitation energy for ECRA clusters. The second remark is that
ECRA clusters with energies higher than \Elb~can be found in some specific
cases (see figure \ref{f:Erelvst90} for the 100 + 100 system) when particles
cannot easily escape from clusters. The evolutions seen for ECRA clusters are
then more likely due to the physics of the collision rather than to the
definition of these clusters.

As in section \ref{s:finaux}, the energy which can be stored in \mod{a 
thermalised} cluster seems
to be limited  almost \mod{throughout} 
the collision time.  The detailed analysis
shows that the exact value of this limitation may depend on reaction time and
on the system size. Before the ``freeze-out'' time, this limit can be higher
than \Elb~if clusters are close enough \mod{together} 
for a long time. This has been observed
for the heaviest systems and for an \Ecm~value close to the  binding energy of
the fused system. After the ``freeze-out'' time, all clusters have an
excitation energy per particle below \Elb.
 
If clusters are excited in a confining "box" (wall, neighboring clusters) which
prevents quick \mod{emission of monomers and small clusters}, 
very high excitation
energies can be deposited in these clusters: they are artificially bound by the
confining wall or the neighboring clusters. If clusters are "free" (no
confining wall, no neighboring clusters), the energy deposition per particle 
can not exceed \Elb.

\section{Conclusions}

The energy deposition in clusters in cluster-cluster collisions  has been
studied in the framework of Classical N Body Dynamics. The mechanism of energy
deposition seems to be the following: the excitation energy of clusters is
mainly due to the relative velocity damping between the projectile and the
target, and to a lesser extent to particle exchanges between them. The
excitation per particle \mod{of thermalised clusters} 
is limited by the energy of the least bound particle in
the cluster. In the low energy regime ($E_{c.m.}/N < E_{LeastBound}$), the
excitation energy increases when the impact parameter decreases and reaches its
maximum value for the most central collisions ($b_{red} \leq 0.5$). For the
intermediate energy regime ($E_{c.m.}/N \approx E_{LeastBound}$) the picture is
almost the same except that for central collisions the incomplete fusion system
has an excitation energy per particle close to  \Elb. For high energy regimes
($E_{c.m.}/N \geq E_{Bind}/N$), the maximum energy deposition is found for
intermediate impact parameters ($b_{red} \approx 0.5$) whereas for central
collisions clusters are less excited and $E*/N$ is always lower than 
$E_{LeastBound}$. 

This limitation of energy deposition is almost verified \mod{throughout} the
collision. High energy deposits  ($E*/N \approx 1.4\>E_{LeastBound}$) have
been found for ECRA clusters and for heavy systems before the ``freeze-out''
time, \mod{which coincides with the thermalisation time of the clusters}. 
This effect is mainly due to the times scales of the reaction. Above the
``freeze-out'' \mod{(thermalisation)} 
time the excitation energy per particle of free clusters is below \Elb. 

This limitation of energy deposition \mod{in thermalised clusters} 
could be an explanation for rather low
excitation energies found in primary \mod{fragments} 
in nucleus-nucleus collisions.
If this limitation is also true for nuclei, this would be in contradiction with
many experimental works in which excitation energies well above the binding
energy have been determined. This could shed new light on caloric curve
analyses and on phase transitions studies.  

\section*{Acknowledgments} 
I would like to warmly thank J.D.Frankland for his
careful reading of this article and for valuable suggestions.


\printfigures

\end{document}